\documentclass[aps,pra,amsmath,twocolumn]{revtex4}

\newcommand{\ket}[1]{\vert #1 \rangle}
\newcommand{\bra}[1]{\langle #1 \vert}
\newcommand{\bracket}[2]{\langle #1 \vert #2 \rangle}

\newcommand{\tr}{\mathop{\mathrm{tr}}\nolimits}
\newcommand{\var}{\textrm{Var}}
\newcommand{\E}{\textrm{E}}
\newcommand{\MSE}{\textrm{MSE}}

\newcommand{\ImPart}{\textrm{Im}}

\newcommand{\vacuum}{0}

\begin{document}

\title{Optimal phase measurements with pure Gaussian states}
\author{Alex Monras}
\affiliation{Grup de F{\'\i}sica Te{\`o}rica \& IFAE, Facultat de
Ci{\`e}ncies, Edifici Cn, Universitat Aut{\`o}noma de Barcelona,
08193 Bellaterra (Barcelona) Spain}
\begin{abstract}
    We analyze the Heisenberg limit on phase estimation for
    Gaussian states. In the analysis, no reference to a phase
    operator is made. We prove that the squeezed vacuum
    state is the most sensitive for a given average photon number.
    We provide two adaptive local measurement schemes that attain the Heisenberg limit asymptotically. One of them
    is described by a positive operator-valued measure and its efficiency is exhaustively
    explored. We also study Gaussian measurement schemes based on
    phase quadrature measurements. We show that homodyne tomography of the appropriate quadrature attains the
    Heisenberg limit for large samples. This proves that this limit can be attained with local projective Von
    Neuman measurements.
\end{abstract}

\maketitle
\section{Introduction}
For a long time quantum phase measurements have been a field of
attention for both theoretical and experimental research in quantum
theory \cite{schleich93, holevo82, holevo84, yuen73, helstrom68,
helstrom76, berry01, berry02, susskind64, shapiro89, shapiro91,
pegg89, braunstein92, smithey93, freyberger94, sanders95, wiseman98,
ou97}. Many schemes have been proposed with the aim of optimally
determining the phase shift produced by a physical process
\cite{sanders95}, usually focusing on quantized harmonic
oscillators. Despite all the efforts, no attempt has ever settled
the issue with full generality \cite{schleich93}. Many different
approaches exist but they often have a restricted range of
applicability. There are various reasons for this situation. The
most important one is the fact that a well defined phase operator
does not exist on the whole Fock space. This seems to indicate that
Heisenberg's uncertainty relation for phase and number may not be
always taken at ``face value" and a rederivation without the use of
such a phase operator is called for (see~\cite{caves95} and
references therein).

In this paper we are concerned with phase measurements on pure
Gaussian states. First, we review the derivation of the Heisenberg
limit using the theory of quantum state estimation
\cite{braunsteincaves94, caves95}, which does not require the
existence of a phase operator. Then we propose a {positive
operator-valued measure} (POVM) that attains the limit by an
adaptive procedure \cite{gillmassar00, hayashi03, matsumoto02}, for
asymptotically large number of copies. We also determine the optimal
Gaussian measurement for phase estimation in this asymptotic regime.

When performing phase measurements on Gaussian states, one usually
implements them on light beams with a given average photon number,
\emph{i.e.,} with a fixed energy. This constraint is always assumed
when comparing different Gaussian states in order to determine their
optimality for phase estimation. The techniques we use, namely,
asymptotics in quantum statistics, are best suited for the problem
at hand, where one usually has a light beam turned on for a large
number of coherence times, which translates into having a large
number of identical copies of such Gaussian state.

\section{Quantum parameter estimation}
\label{sec:estimationth} In this section we review the theory of
quantum parameter estimation \cite{helstrom76, braunsteincaves94,
caves95, gill00, gill04, dariano00, matsumoto02}, which leads to a
useful generalization of the Heisenberg uncertainty relation.

Let us assume we have a family of pure states $\rho(\theta)$
labelled by a parameter $\theta$. In our case the parameter is the
phase. This family corresponds to a curve in Hilbert space, and
translations along such curve are generated by a hermitian operator
$G$ in the usual manner. In our case the generator is the number
operator $G=a^\dagger a$:
\begin{equation}
    \rho(\theta)=\exp(-i\theta G)\rho(0)\exp(i\theta G).
\end{equation}

Throughout this paper we use the following notation. We will write
$\ket n$ to denote Fock states, \emph{i.e.}, states with well
defined number. With $\mathcal U_\theta$ we denote the unitary
operator that yields the state $\rho(\theta)$ from the vacuum,
namely
\begin{equation}
    \rho(\theta)=\mathcal U_\theta \ket0\bra0\mathcal U^\dagger_\theta.
\end{equation}
We define the vectors $\ket{\phi_n}$ as the Fock states transformed
by $\mathcal U_\theta$,
\begin{equation}
    \ket{\phi_n}=\mathcal U_\theta\ket n;
\end{equation}
hence, $\rho(\theta)=\ket{\phi_0}\bra{\phi_0}$.

Our aim is to determine the parameter $\theta$ by performing the
best possible POVM measurement on the system. For this, we propose
an estimator $\hat\theta(x)$, which is a function of the possible
outcomes~$x$ of the measurement. Hence, the expectation value of
such estimator is \cite{cramer46}
\begin{equation}
    \E_\theta[\hat\theta]=\sum_{x}p(x|\theta) \hat\theta(x),
\end{equation}
and its variance
\begin{equation}
    \var_\theta[\hat\theta]=\sum_{x}p(x|\theta)(\hat\theta(x)-\E_\theta[\hat\theta])^2,
\end{equation}
where the subscript $\theta$ means that the expectation value is
taken with probability distribution $p(x|\theta)$. In the quantum
case, they are given by the Born rule
$p(x|\theta)=\tr[\rho(\theta)E_x]$, where $\{E_x\}$ are the elements
of a POVM.

An estimator is called {unbiased} when
\begin{equation}
    \E_\theta[\hat\theta]=\theta.
\end{equation}
In this case, the variance is equivalent to the mean squared error
(MSE):
\begin{equation}
    \MSE[\hat\theta]\equiv\E[(\hat\theta-\theta)^2]=\var[\hat\theta]+(\E[\hat\theta]-\theta)^2.
\end{equation}
A well known theorem in statistics gives a lower bound to the
variance of any unbiased estimator. It is the so-called Cram\'er-Rao
bound, \cite{cramer46}
\begin{equation}
    \label{eq:CramerRao}
    \var_\theta[\hat\theta]\geq\frac{1}{F(\theta)},
\end{equation}
where $F(\theta)$ is the Fisher information associated with the
measurement,
\begin{equation}
    \label{eq:fisher}
    F(\theta)=\sum_{x\in X_+}p(x|\theta)\left[\frac{\partial \log p(x|\theta)}{\partial
    \theta}\right]^2,
\end{equation}
and the sum runs over the set of \textit{possible} outcomes~$X_+$,
i.e. those with $p(x|\theta)\neq0$.

Furthermore, the Braunstein-Caves inequality
\cite{braunsteincaves94} sets an upper bound on the Fisher
information:
\begin{equation}
    \label{eq:BraunsteinCaves}
    F(\theta)\leq H(\theta),
\end{equation}
where $H(\theta)$ does not depend on the specific measurement being
performed. $H(\theta)$ is sometimes regarded as the Quantum Fisher
Information (QFI). It is defined as \cite{helstrom76,
braunsteincaves94, gillmassar00, matsumoto02}
\begin{equation}
    \label{eq:QFI}
    H(\theta)=\tr[\rho(\theta)\lambda(\theta)^2],
\end{equation}
where $\lambda$ is the {symmetric logarithmic derivative} (SLD),
defined as the hermitian operator that fulfills
\begin{eqnarray}
    \nonumber
    \frac{\partial \rho(\theta)}{\partial \theta}&=&-i[G,\rho(\theta)]=\\
    &=&\frac{1}{2}\left[\lambda(\theta)\rho(\theta)+\rho(\theta)\lambda(\theta)\right].
\end{eqnarray}

Eqs.~\eqref{eq:CramerRao} and \eqref{eq:BraunsteinCaves} set a
fundamental bound on the variance of any unbiased estimator. This
bound only depends on the geometrical properties of the curve
$\rho(\theta)$. In our case, the SLD is
\begin{equation}
    \lambda(\theta)=2i\left(\ket{\phi_0}\bra\psi-\ket\psi\bra{\phi_0}\right)=-2i[G,\rho(\theta)],
\end{equation}
where
\begin{equation}
    \label{eq:psi}
    \ket\psi=\left(1-\ket{\phi_0}\bra{\phi_0}\right)G\ket{\phi_0}.
\end{equation}
The eigenvectors of $\lambda$ are $\ket{\phi^\pm}=\ket\psi\pm
i\sqrt{\bracket{\psi}{\psi}}\ket{\phi_0}$, with the corresponding
eigenvalues given by $l_\pm=\pm2\sqrt{\bracket{\psi}{\psi}}$. Note
that $\bracket\psi{\phi_0}=0$. Working out the value of $H$ we get
\begin{equation}
    H(\theta)=4\bracket{\psi}{\psi}=4\langle\Delta G\rangle^2_\theta,
\end{equation}
where $\langle\Delta G\rangle_\theta^2$ is defined, as usual, as
$\bra{\phi_0}G^2\ket{\phi_0}-\bra{\phi_0}G\ket{\phi_0}^2$. Thus,
\begin{equation}
    \label{eq:newHeisenberg}
    \var_\theta[\hat\theta]\langle\Delta G\rangle_\theta^2\geq\frac{1}{4}.
\end{equation}
This expression is similar to Heisenberg's uncertainty relation for
canonically conjugate variables, but has some advantages (see
\cite{caves95}). First of all, it has been derived without the use
of any phase operator. In fact, the only operator we need is the
phase shift generator, \emph{i.e.}, the number operator. On the
other hand, it sets a lower bound on the variance of an estimator,
whereas the standard uncertainty relation does not concern
optimality but only variances obtained from measurements of
observables (\emph{i.e.}, self-adjoint operators). In this sense,
 Eq.~\eqref{eq:newHeisenberg} is more general.

Now assume we have $N$ identical copies of the same unknown state
$\rho(\theta)$. In this case the collective state
$\rho(\theta)^{\otimes N}$ is still a member of a one-parameter
family, and it is straightforward to show that the corresponding QFI
scales as $N$. Thus combining Eqs. \eqref{eq:CramerRao} and
\eqref{eq:BraunsteinCaves}, with $H$ replaced by the $N$-copy QFI
and $\hat\theta$ denoting any unbiased estimator based on any
measurement (collective or individual) on the $N$ copies, we have
%
\begin{equation}
    \label{eq:NQCRB}
    \var_{N,\theta}[\hat\theta]\langle\Delta
    G\rangle_\theta^2\geq\frac{1}{4N},
\end{equation}
where the subscript $N$ stands for the number copies.

An important issue is the attainability of these bounds. The
Braunstein-Caves inequality is known to be saturable
\cite{braunsteincaves94, gill00}, \emph{i.e.}, there exists a POVM
that gives the equality in \eqref{eq:BraunsteinCaves} [it is given
by $\lambda(\theta)$, see Sec. \ref{sec:optimalPOVM} for details].
Note, however, that in general $\lambda(\theta)$ is not constant and
the optimal POVM depends on the true value of $\theta$. This
suggests that, in order to attain the equality in
\eqref{eq:BraunsteinCaves} for large $N$, one needs an adaptive
scheme. The measurement at a given step of the estimation scheme may
need to be optimized using the data gathered from the previous
steps. The Cram\'er-Rao bound is also known to be asymptotically
saturable by the so-called {maximum likelihood estimator} (MLE).
That is, for multiple identical measurements, the MLE has a variance
which approaches the inverse of the Fisher information, thus
attaining equality also in \eqref{eq:CramerRao}. All this seems to
indicate that the optimal scheme should consist of at least two
steps: A preliminary rough estimate of $\theta$, using a vanishing
fraction of copies ($N^\alpha$, with $0<\alpha<1$), and a refinement
using the remaining $\bar N=N-N^\alpha$ copies, where all
measurements are identical in the last step.

Therefore, if $N\rightarrow\infty$ one can reasonably expect that
\begin{equation}
    \lim_{N\rightarrow\infty}N~\var_{\bar N,\theta}[\hat\theta]=\frac{1}{4\langle\Delta G\rangle_\theta^2}
\end{equation}
for the MLE, provided the optimal POVM measurement is performed in
step two. Here the subscript $\bar N$ indicates that the variance is
obtained from the outcomes of the last $\bar N$ identical
measurements. This will be loosely written as
\begin{equation}
    \label{eq:AssNCRB}
    \var_\theta[\hat\theta]\sim\frac{1}{4\langle\Delta G\rangle_\theta^2}.
\end{equation}
We show below that \eqref{eq:AssNCRB} holds if the number of copies
used in each of the two steps of the estimation scheme (\emph{i.e.},
$\alpha$) is appropriately chosen.

A word should be said about asymptotically unbiased estimators.
These are estimators whose expectation values converge to the
parameter in the large sample limit, but have some vanishing bias,
\emph{i.e.},
\begin{equation}
    \label{eq:AssUnbiased}
    \E[\hat\theta]=\theta+O(N^{-\alpha}),
\end{equation}
with $\alpha>0$. In this case the correct quantifier of the
sensitivity is the MSE. If $\alpha>1/2$ the MSE is asymptotically
equivalent to the variance, since
\begin{equation}
    \MSE[\hat\theta]=\var[\hat\theta]+O(N^{-2\alpha}),
\end{equation}
and we note that the leading order contribution in inverse powers of
$N$ is solely given by $\var[\hat\theta]$.
\section{Optimal Gaussian state}
Next we determine the optimal Gaussian state, which provides the
highest sensitivity to phase measurements within our approach,
\emph{i.e.}, the one that maximizes its QFI. The state is obtained
by sequentially applying a series of operations on the vacuum:
(\emph{i})~a squeezing along a fixed direction (say $Q$),
$S(r)=\exp[(r/2)(a^2-a^\dagger{}^2)]$; (\emph{ii})~a displacement,
$D(\alpha)=\exp\left(\alpha a^\dagger-\alpha^*a\right)$ with
real~$\alpha$; (\emph{iii})~the unknown phase shift,
$U(\theta)=\exp\left(-i\theta a^\dagger a\right)$. The state after
applying these operations is
\begin{equation}
    \label{eq:state}
    \ket{\phi_0}=U(\theta)D(\alpha)S(r)\ket\vacuum,
\end{equation}
and its QFI is given by
\begin{eqnarray}
    H(\theta)&=&4\langle\Delta a^\dagger a\rangle^2_\theta\\
    \nonumber
    &=&4\left[|\alpha|^2\left(\cosh r-\sinh
    r\right)^2+2\sinh^2r\cosh^2r\right].
\end{eqnarray}
Notice that actually $H$ does not depend on the phase. The energy of
the state \eqref{eq:state} is
\begin{equation}
    \langle n\rangle_\theta=\langle a^\dagger
    a\rangle_\theta=|\alpha|^2+\sinh^2r.
\end{equation}
We aim at maximizing $H$ with the constraint to have a fixed average
number of photons. Using Lagrange multipliers it is straightforward
to find that the most sensitive choice is $\alpha=0$. As intuition
dictates, the highest sensitivity is achieved by employing all
available energy in squeezing the vacuum. Therefore, the optimal
gaussian state is $\ket{\phi_0}=U(\theta)S(r)\ket\vacuum$, for which
\begin{equation}
    \label{eq:MaxQFI}
    H=\cosh 4r-1=8(\langle n\rangle^2+\langle n\rangle).
\end{equation}
Hence, optimal phase estimation has a variance that goes as
\begin{equation}
    \label{eq:OptimalPerformance}
    \var[\hat\theta]\sim\frac{1}{8(\langle n\rangle^2+\langle
    n\rangle)}.
\end{equation}
\section{Asymptotically optimal few-outcome measurement}
\label{sec:optimalPOVM} Following the derivation of the
Braunstein-Caves inequality \cite{braunsteincaves94}, one can check
that the optimality conditions on the POVM elements $\{E_x\}$ are
(see \cite{gill00})
\begin{eqnarray}
    \label{eq:optimalPOVM1}
    \ImPart\{\tr[\rho(\theta)E_x\lambda(\theta)]\}&=&0,\\
    \nonumber
    \sqrt{\rho(\theta)}\lambda(\theta)\sqrt{E_x}&=&k_x\sqrt{\rho(\theta)}\sqrt{E_x},
\end{eqnarray}
for all $x$ that have non-vanishing probabilities, where $k_x$ are
some constants. Assuming pure states $\rho^2=\rho$, and the fact
that any POVM can be decomposed into rank one operators (so that
$\sqrt{E_x}\propto E_x$), one may write the second optimality
condition as
\begin{eqnarray}
        \label{eq:optimalPOVM2}
    \rho(\theta)\lambda(\theta)E_x&=&k_x\rho(\theta)E_x.
\end{eqnarray}
Finally, one can check that a sufficient condition for
Eqs.~\eqref{eq:optimalPOVM1} and \eqref{eq:optimalPOVM2} to hold is
that $\{E_x\}$ project onto the eigenspaces of the SLD. Therefore,
an optimal POVM can be chosen to have three elements,
$\{E_+,E_-,E_0\}$, given by
\begin{eqnarray}
    E_\pm=\frac1{2\bracket{\psi}{\psi}}\ket{\phi^{\pm}}\bra{\phi^{\pm}}
\end{eqnarray}
and
\begin{equation}
    E_0=\openone-E_+-E_-.
\end{equation}
For completeness we give the explicit form of the
states~$\ket{\phi^{\pm}}$. They are
\begin{equation}
    \ket{\phi^\pm}=\sqrt{2}\sinh r\cosh r\left(\pm i\ket{\phi_0}-\ket{\phi_2}\right),
\end{equation}
which, after normalizing become
\begin{equation}
    \frac{\ket{\phi^\pm}}{{\big|\ket{\phi^\pm}}\big|}=\frac{\pm i\ket{\phi_0}-\ket{\phi_2}}{\sqrt
    2}=\mathcal U_\theta\frac{\pm i\ket 0-\ket2}{\sqrt
    2}.
\end{equation}
This measurement resembles tomography applied to estimation of spin
states lying close to the $z$ axis of the Bloch sphere, where spin
measurements along the $x$ and $y$ directions are optimal
\cite{gillmassar00, bagan05}.

From the hermiticity of $\lambda$ it is straightforward to check
that $\{E_+,E_-,E_0\}$ are positive and, furthermore, that they
represent projective von Neuman measurements. The outcomes of these
measurements will be used in the maximum likelihood analysis of the
next section, where its corresponding expectation value and MSE will
be discussed.

\section{Maximum likelihood estimation}
\label{sec:MLE}
Assume one has performed step one, a series of
non-optimal measurements on a vanishingly small number of copies,
$N^\alpha$, of our optimal Gaussian state $\ket{\phi_0}$, and has
obtained a rough guess $\hat\theta_0$ of its phase. Let us write
$\hat\theta_0=\theta-\delta \theta$, where $\delta \theta$
represents the error in this first step, which is assumed to be
small. Let $\ket{\hat{\phi}_0}$ be the guessed state. We design
(step two) the optimal measurement on the remaining $\bar
N=N-N^\alpha$ copies as if $\hat\theta_0$ were the true phase.

Let $N_{\pm,0}$ be the number of times the outcomes $\pm,0$ are
obtained. The likelihood function for $\theta$ is \cite{cramer46}

\begin{equation}
    \mathcal L(\theta)=\frac{\bar
    N!}{N_+!N_-!N_0!}p(+|\theta)^{N_+}p(-|\theta)^{N_-}p(0|\theta)^{N_0}.
\end{equation}
Numerical maximization of this function yields the strict~MLE.
Recall that this is known to attain the Cram\'er-Rao bound in the
limit of large number of copies. We can, however, analytically
compute an approximate estimator, which converges to the MLE in the
large sample limit and has the same nice properties, \emph{i.e.}, it
is asymptotically unbiased and attains the Cram\'er-Rao bound
asymptotically. These properties will be checked explicitly.

Taylor expanding the true state around the first-step guess we have
\begin{eqnarray}
    \rho(\theta)\!=\!\rho(\hat\theta_0)\!+\!\frac{1}{2}\!\left[\!\lambda(\hat\theta_0)\rho(\hat\theta_0)\!+\!\rho(\hat\theta_0)\lambda(\hat\theta_0)\right]\!\!\delta\theta\!+\!O(\delta\theta^2\!),
\end{eqnarray}
and the probabilities of obtaining the outcomes $x=\pm,0$ are
\begin{eqnarray}
    p(\pm|\theta)&=&\frac{1}{2}\pm\sqrt{\bracket{\hat\psi}{\hat\psi}}\delta\theta+O(\delta\theta^2),\\
    p(0|\theta)&=&O(\delta\theta^2),
\end{eqnarray}
where $\ket{\hat \psi}$ is given by Eq. \eqref{eq:psi} by replacing
$\ket{\phi_0}$ for $\ket{\hat\phi_0}$. With this,
$\bracket{\hat\psi}{\hat\psi}=\langle\Delta n\rangle^2$. Thus,
\begin{eqnarray}
    \displaystyle
    &&\kern-2em\frac{\partial \log\mathcal L}{\partial \theta}=\sum_{x\in\{+,-,0\}}\frac{N_x}{p(x|\theta)}\frac{\partial p(x|\theta)}{\partial
    \theta}\\
    \nonumber
    &&\kern-2em=4\langle\Delta n\rangle\left[\frac{N_+-N_-}{2}-(N_++N_-)\langle\Delta
    n\rangle\delta\theta\right]+O(\delta\theta^2).
\end{eqnarray}
Equating this to zero yields the approximate MLE at leading order in
$\delta\theta$:
\begin{equation}
    \hat\theta_{\textrm{MLE}}\equiv\hat\theta_0+\frac{N_+-N_-}{2N_{\textrm{inf}}\langle\Delta n\rangle}.
\end{equation}
Here $N_{\textrm{inf}}=N_++N_-$ is the total number of
``informative" outcomes. This estimator is based on disregarding the
improbable and noninformative outcomes $x=0$. Our approximate MLE is
undefined in the rare event that $N_{\textrm{inf}}=0$. In this case
we may just keep the preliminary estimate and define
$\hat\theta_{\textrm{MLE}}\equiv \hat\theta_0$.

Disregarding the unlikely $x=0$ outcomes, the estimator is just
based on a binomial distribution, and for a fixed value of
$N_{\textrm{inf}}$ we have
\begin{eqnarray}
    \nonumber
    \E_{\theta,N_{\textrm{inf}}}[\hat\theta_{\textrm{MLE}}]&=&\sum_{N_+=0}^{N_{\textrm{inf}}}\hat\theta_{\textrm{MLE}}\binom{N_{\textrm{inf}}}{N_+}q^{N_+}(1-q)^{N_-}\\
    &=&\hat\theta_0+\frac{2q-1}{2\langle \Delta n\rangle},
\end{eqnarray}
where $q$ is, to the relevant order, $q=1/2+\langle \Delta
n\rangle\delta \theta+O(\delta\theta^2)$. This yields
\begin{equation}
    \E_{\theta,N_{\textrm{inf}}}[\hat\theta_{\textrm{MLE}}]=\theta+O(\delta\theta^2).
\end{equation}
Since $\delta\theta^2=O(N^{-\alpha})$, the expectation value is
asymptotically unbiased [recall Eq. \eqref{eq:AssUnbiased}]. It is
also straightforward to compute the MSE of this estimator, for a
given number $N_{\textrm{inf}}$ of informative outcomes :
\begin{multline}
    \MSE_{\theta,N_{\textrm{inf}}}[\hat\theta_{\textrm{MLE}}]\\
    =\frac{1}{4\langle\Delta
    n\rangle^2N_{\textrm{inf}}}+O\left(\frac{\delta\theta^2}{\bar
    N}\right)+O(\delta\theta^4),
\end{multline}
for $N_{\textrm{inf}}>0$, and
\begin{equation}
    \MSE_{\theta,0}[\hat\theta_{\textrm{MLE}}]=\delta\theta^2,
\end{equation}
for $N_{\textrm{inf}}=0$.

To compute the full MSE of our scheme we just need to average over
all possible values of $N_{\textrm{inf}}$. This yields (see Appendix
\ref{sec:appAverage} for details)
\begin{eqnarray}
    \MSE_\theta[\hat\theta_{\textrm{MLE}}]&=&\sum_{N_{\textrm{inf}}=0}^{\bar N}\MSE_{\theta,N_{\textrm{inf}}}[\hat\theta_{\textrm{MLE}}]\\
    \nonumber
    &&\times\binom{\bar N}{N_{\textrm{inf}}}p^{N_{\textrm{inf}}}(1-p)^{\bar N-N_{\textrm{inf}}}\\
    &=&\frac{1}{\bar N}\left[\frac{1}{4\langle\Delta
    n\rangle^2}+O(\delta\theta^2)\right],
\end{eqnarray}
where $p$ is the probability of getting an informative outcome in a
given measurement [$p=1-O(\delta\theta^2)$]. Clearly enough, by
choosing $\alpha>1/2$ the error of the first step only contributes
to subleading orders in the variance. If, e.g., $\alpha=2/3$ one
sees that in the large $N$ limit, the~$\MSE[\hat\theta]$ reduces to
\begin{equation}
    \MSE_\theta[\hat\theta_{\textrm{MLE}}]=\frac{1}{8(\langle n\rangle^2+\langle n\rangle)~N}+O\left(\frac{1}{N^{4/3}}\right).
\end{equation}
Hence, the optimal performance displayed
in~\eqref{eq:OptimalPerformance}~is asymptotically attained.

\section{Asymptotically optimal Gaussian measurements}
Now we face the problem of determining the minimal phase variance
that can be attained by means of dyne measurements. These consist of
the simultaneous measurement of two conjugate variables, such as $Q$
and $P$, or their phase generalizations
$U(\theta')QU^\dagger(\theta')$ and $U(\theta')PU^\dagger(\theta')
$. They are typically performed by splitting the signal state. Often
this is done by means of a beam splitter, the so-called eight-port
homodyne detector \cite{leonhardt97}. Another possibility is
heterodyne detection. In any case, the Arthurs \& Kelly theorem
(see, e.g., \cite{leonhardt97}) guarantees that further noise will
appear in the detection process. In the eight-port homodyne detector
it appears as vacuum fluctuations entering through the unused port
of the beamsplitter, which is sometimes called \emph{auxiliary
port}. By introducing some sort of squeezed vacuum in the auxiliary
port, one can reduce the noise in one quadrature, at the expense of
increasing the noise in the corresponding conjugate quadrature.
These measurements can be mathematically expressed as a covariant
measurement \cite{holevo82, holevo84} with POVM elements given by
\begin{eqnarray}
    \label{eq:POVM}
    E(\chi)=\frac{1}{2\pi}D(\chi)\sigma_0D^\dagger(\chi)=\frac{1}{2\pi}\sigma_{\chi},
\end{eqnarray}
where $(q,p)\equiv\chi$ are the outcomes and represent coordinates
in phase space, and $D(\chi)$ is the displacement operator
$D(\chi)=\exp i(pQ-qP)$. The density matrix $\sigma_0$ is the
ancilla entering through the auxiliary port of the beamsplitter.
From the covariance of the measurement, $\sigma_0$ can be taken to
be a squeezed vacuum state ($\langle q\rangle=\langle p\rangle=0$)
without loss of generality. One can relate the probability of
obtaining the outcome $\chi$, $dp(\chi|\theta)$, with the fidelity
\cite{uhlmann76}, $\mathcal
F(\sigma_\chi,\rho(\theta))=\tr\sqrt{\rho(\theta)^{1/2}\sigma_\chi\rho(\theta)^{1/2}}$,
as
\begin{eqnarray}
    \nonumber
    dp(\chi|\theta)&=&\tr[\rho(\theta)E(\chi)]d^2\chi\\
    \label{eq:dProb}
    &=&\frac{d^2\chi}{2\pi}\mathcal F(\sigma_\chi,\rho(\theta)).
\end{eqnarray}
When $\sigma_0$ is a Gaussian state the measurement is said to be
Gaussian and the outcomes are Gaussian distributed. Moreover, when
one of the two states is pure, the fidelity can be easily expressed
as $\mathcal F(\sigma_\chi,\rho(\theta))^2=2\sqrt{\det
M(\theta)}e^{-\chi^t M(\theta)\chi}$ \cite{wolf05}, where
$M(\theta)=(\gamma_0+\gamma_\theta)^{-1}$. The covariance matrices
of the input and the auxiliary states are
\begin{eqnarray}
    \gamma_\theta&=&
    R^t(\theta)~S~R(\theta),\\
    \gamma_0&=&R^t(\theta')~T~R(\theta'),
\end{eqnarray}
with
\begin{eqnarray}
    &R(\theta)=
    \left(
        \begin{array}{cc}
            \cos \theta&-\sin\theta\\
            \sin\theta&\cos\theta
        \end{array}
    \right),&\\
\nonumber  \\
    \label{eq:covariances}
    &S=\left(
        \begin{array}{cc}
            s^2&0\\
            0&1/s^2
        \end{array}
    \right),\qquad T=\left(
        \begin{array}{cc}
            t^2&0\\
            0&1/t^2
        \end{array}
    \right),&
\end{eqnarray}
where $s\equiv e^{-r}$ and $\theta$ ($t\equiv e^{-r'}$ and
$\theta'$) are the squeezing parameter and phase of the input state
(ancilla). Recall that the diagonal matrix elements of the
covariance matrices give the variances of the canonical observables
$\langle\Delta Q\rangle^2=[\gamma_\theta]_{11}/2$ and $\langle\Delta
P\rangle^2=[\gamma_\theta]_{22}/2$.

From \eqref{eq:dProb} the probability density is
\begin{eqnarray}
    \label{eq:Prob}
    p(\chi|\theta)=\frac{\sqrt{\det M(\theta)}}\pi\exp[-\chi^t
    M(\theta)\chi].
\end{eqnarray}
The Fisher information of such measurement can be readily computed
from \eqref{eq:fisher}:
\begin{equation}
    \label{eq:symbolicfisher}
    F(\theta)=\frac{1}{2}\tr[M'M^{-1}M'M^{-1}],
\end{equation}
where $M'$ denotes the derivative of $M(\theta)$ w.r.t.~$\theta$ and
$M^{-1}$ its inverse. The $\theta$ dependency has been dropped for
readability reasons. Further manipulations show that the Fisher
information can be written as
\begin{equation}
    F(\theta)=\frac{1}{2}\tr[\Gamma^{-1}\Sigma\Gamma^{-1}\Sigma],
\end{equation}
with
\begin{eqnarray}
\nonumber
    \Gamma&=&S+R^t(\theta'-\theta)~T~R(\theta'-\theta)\\
\nonumber
    &=&(\cosh 2r+\cosh2r')\openone\\
\nonumber
    &&-(\sinh2r+\sinh2r'\cos\varphi){\boldsymbol\sigma}_3\\
    &&+\sinh2r'\sin\varphi~{\boldsymbol\sigma}_1
\end{eqnarray}
and
\begin{eqnarray}
    \Sigma=R^t(\pi/2)~S+S~R(\pi/2)=2\sinh2r~{\boldsymbol\sigma}_1,
\end{eqnarray}
where, as expected, $\varphi=2(\theta'-\theta)$ is the only relevant
angular parameter, and ${\boldsymbol\sigma_i}$ ($i=1,2,3$) are the
standard Pauli matrices. With this, $F(\theta)$ reads
\begin{eqnarray}
    \nonumber
    F(\theta)&=&\frac{2\sinh^22r}{(\cosh2r\cosh2r'-\sinh2r\sinh2r'\cos\varphi+1)^2}\\
    \nonumber
    &&\times(\cosh2r\cosh2r'-\sinh2r\sinh2r'\cos\varphi\\
    \label{eq:Fisher}
    &&+\sinh^22r'\sin^2\varphi+1).
\end{eqnarray}
Here the dependence of $\varphi$ on $\theta$ is implicit. To
optimize the measurement, we find the values of $\theta'$ and $r'$
that maximize the Fisher information. Imposing $\partial F/\partial
\varphi=0$ we find three possible extremal points: $\varphi=0$ and
$\varphi=\pm\varphi_0(s,t)$ [see Eq.~\eqref{eq:optimalVarphi} below
for the expression of~$\varphi_0$]. We will drop the arguments of
$\varphi_0$ where no confusion arises. The angle $\varphi=0$ is a
maximum if the squeezing in the auxilliary port of the beamsplitter
is weaker than a threshold, $1<t<t_{\textrm{thr}}(s)$, given by
\begin{equation}
    t_{\textrm{thr}}(s)=\frac{1}{2s}\sqrt{s^4-1+\sqrt{s^8+14s^4+1}}.
\end{equation}
In this case, this is the only extremal point of $F(\theta)$. When
$t>t_{\textrm{thr}}$ the trivial solution $\varphi=0$ becomes a
minimum and the solutions $\varphi=\pm\varphi_0$ become maxima.

The maximum Fisher information below the threshold
($1<t<t_{\textrm{thr}}$) is
\begin{equation}
    F(\theta')=\frac{t^2(1-s^4)^2}{s^2(s^2+t^2)^2}=\frac{\sinh^22r}{\cosh^2(r-r')}.
\end{equation}
As a function of $t$, the maximum of $F(\theta)$ is at the critical
point $t=t_{\textrm{thr}}$.

Over the threshold ($t>t_{\textrm{thr}}$) the optimal angle is given
by
\begin{equation}
    \label{eq:optimalVarphi}
    \cos\varphi_0=\frac{2\cosh4r'\sinh2r+\cosh2r'\sinh4r}{(3+\cosh4r)\sinh2r'+2\cosh2r\sinh4r'},
\end{equation}
$\varphi_0\geq0$ [recall the definition of $s$ and $t$ right below
Eq.~\eqref{eq:covariances}]. According to the definition of
$\varphi$, we have $\theta'=\theta\pm\varphi_0/2$. The maximum
Fisher information is
\begin{multline}
    F\left(\theta'\mp\varphi_0/2\right)=\sinh^22r\\
    \times\frac{3+\cosh4r+8\cosh2r\cosh2r'+4\cosh4r'}{4(\cosh2r+\cosh2r')^2}.
\end{multline}
As $t\rightarrow\infty$ ($r'\rightarrow-\infty$), the optimal angle
becomes
\begin{eqnarray}
    \label{eq:optimalPhi}
    \Phi(s)&=&\lim_{t\rightarrow\infty}\varphi_0(s,t)\\
    \nonumber
    &=&\arccos\left(\frac{s^4-1}{s^4+1}\right)=\arccos(-\tanh2r),
\end{eqnarray}
and the Fisher information approaches the value
\begin{equation}
    \label{eq:Qfisher}
    \lim_{t\rightarrow\infty}F\left(\theta'\mp\varphi_0/2\right)=\frac{(1-s^4)^2}{2s^4}=\cosh4r-1.
\end{equation}
This is precisely the Heisenberg limit found in
Sec.~\ref{sec:estimationth} [see also Eqs. \eqref{eq:MaxQFI} and
\eqref{eq:OptimalPerformance}]. Hence, optimal phase sensitivity is
attained with a POVM measurement given by Eq.~\eqref{eq:POVM} and an
auxiliary state $\sigma_0$ characterized by $t\rightarrow\infty$ and
$\theta'=\theta\pm\Phi/2$. This is an infinitely squeezed state
phase-shifted w.r.t.~the true phase. In Sec. \ref{sec:Estimation} we
present a two-step adaptive implementation of this measurement which
can be derived proceeding along the same lines as in Sec.
\ref{sec:MLE}. This scheme also attains  the Heisenberg limit.

\section{Interpretation of the optimal gaussian measurement}
In this section we provide an interpretation of this apparently
unphysical proposal of a Gaussian measurement with infinite
squeezing in the auxiliary port. Introducing an infinitely squeezed
ancilla in this port is equivalent to having zero noise in one
quadrature, say $P'$, at the expense of introducing infinite noise
in the correspondingly conjugate one, $Q'$. This makes the readings
of $Q'$ to be just random noise with no information about the signal
state. Therefore, one has only a measurement of~$P'$. This strongly
supports the idea that, after all, only homodyne detection is needed
to implement the measurement under discussion. Below we provide a
mathematical justification of this interpretation, and show that the
marginal POVM elements, \emph{i.e.}, those obtained by integrating
Eq.~\eqref{eq:POVM} over $q'$, become exactly the rank-one
projectors of the $P'$ measurement.

Let us start by giving the precise definition of the observables
$Q'$ and $P'$:
\begin{eqnarray}
    \label{eq:QandP}
    Q'&=&U(\theta')Q~U^\dagger(\theta')=Q\cos\theta'-P\sin\theta',\\
    P'&=&U(\theta')P~U^\dagger(\theta')=Q\sin\theta'+P\cos\theta'.
\end{eqnarray}
This is just a rotation in phase space, thus the coordinates
transform in the same manner
\begin{equation}
    \chi'=R(\theta')\chi,
\end{equation}
and the displacement operators $D(\chi)$ transform as
\begin{equation}
    \label{eq:newD}
    D'(\chi')\equiv U(\theta')D(\chi')U^\dagger(\theta')=D(\chi).
\end{equation}
Let $\ket{Q;q}$ and $\ket{Q';q}$ be the eigenstates of $Q$ and $Q'$
respectively, where $q$ is the eigenvalue, so that
$\ket{Q';q}=U(\theta')\ket{Q;q}$. Since $\sigma_0$ is
\begin{equation}
    \sigma_0=U(\theta')S(r')\ket0 \bra
    0S^\dagger(r')U^\dagger(\theta'),
\end{equation}
the POVM elements of Eq. \eqref{eq:POVM} can be cast as
\begin{eqnarray}
    \nonumber
    E(\chi)&=&\frac{1}{2\pi}U(\theta')D(\chi')S(r')\ket0\bra0S^\dagger(r')D^\dagger(\chi')U^\dagger(\theta')\\
    \nonumber
    &=&\int\frac{dq_1dq_2}{2\pi}~
    \ket{Q';q_1}\bra{Q';q_2}\\
    &&\times e^{ip'(q_1-q_2)}\psi(q_1-q')\psi^*(q_2-q'),
\end{eqnarray}
where we have introduced the squeezed-vacuum wave function
$\psi(q)$, defined as $\psi(q)=\bra{Q;q}S(r')\ket 0$.

We can compute the marginal POVM elements $F(p')$ by integrating
$E(\chi)$ over $dq'$,
\begin{eqnarray}
    \nonumber
    F(p')&=&\int dq'E(\chi)=\int dq'E(R^{-1}(\theta')\chi')\\
    \nonumber
    &=&\int\frac{dq_1dq_2}{2\pi}~
    \ket{Q';q_1}\bra{Q';q_2}\\
    &\times&e^{ip'(q_1-q_2)}\int dq'\psi(q_1-q')\psi^*(q_2-q').
\end{eqnarray}
Performing the $dq'$ integral yields
\begin{equation}
    \label{eq:integral}
    \int
    dq'\psi(q_1-q')\psi^*(q_2-q')=\exp\left[-\frac{e^{2r'}}{4}(q_1-q_2)^2\right],
\end{equation}
where we have used that \cite{bagan06}
\begin{equation}
    \psi(q)=\frac{e^{r'/2}}{\pi^{1/4}}\exp\left(-\frac{e^{2r'}}{2}q^2\right).
\end{equation}
With this, the marginal POVM reads
\begin{eqnarray}
    \nonumber
    F(p')&=&\int \frac{dq_1dq_2}{2\pi}\ket{Q';q_1}\bra{Q';q_2}\\
    &\times&\exp\left[ip'(q_1-q_2)-\frac{e^{2r'}}{4}(q_1-q_2)^2\right],
\end{eqnarray}
which in the limit $r'\rightarrow-\infty$ converges to
\begin{equation}
    F(p')=\ket{P';p'}\bra{P';p'}.
\end{equation}

 By rotating the canonical operators, $Q'$ and $P'$ we have diagonalized the covariance matrix of $\sigma_0$,
and got rid of the correlations between $Q$ and $P$. We thus put all
the noise introduced by the auxiliary state into the~$Q'$
quadrature, and put all the information of the signal state into the
$P'$ quadrature. This enables us to use the marginal POVMs without
loss of information, which in turn means that it suffices to measure
the $P'$ quadrature, \emph{i.e.}, the informative one.

Another interpretation follows from \cite{leonhardt97}, where it is
shown that introducing a squeezed state in the auxiliary port is
equivalent to leaving the vacuum and tuning the beamsplitter to a
transmittance/reflectance other than 50\%/50\%. The limit of
infinite squeezing amounts to having a completely transmitting or
reflecting mirror, which again converts heterodyning into
homodyning.

\section{Estimation Scheme}
\label{sec:Estimation} The analysis carried out so far tells us that
optimal phase measurements can be performed by means of homodyne
tomography of the appropriate quadrature
$\theta'=\hat\theta_0\pm\Phi(s)/2$, where $\Phi(s)$ is given by
Eq.~\eqref{eq:optimalPhi} and~$\hat\theta_0$ is the estimate of the
first step measurement, which is not assumed to be optimal. Let us
call this quadrature~$P'$. The outcomes will be gaussian distributed
with zero mean (recall that our signal state is a squeezed vacuum)
and variance given by
\begin{equation}
    \sigma^2(\theta',\theta)=\frac{1+s^4+(1-s^4)\cos2(\theta'-\theta)}{4s^2},
\end{equation}
as can be seen by diagonalizing the $M$ matrix in
Eq.~\eqref{eq:Prob} through the transformation
$\chi'=R(\theta')\chi$.

It is straightforward to check that the Fisher information provided
by this distribution is exactly that given by
Eq.~\eqref{eq:Qfisher}.

The maximum likelihood analysis of the data yields the condition
$\sigma^2(\theta',\hat\theta_{\textrm{MLE}})=\sum p_i^2/{\bar N}$,
where $\{p_1,\ldots,p_{\bar N}\}$ is the set of outcomes
corresponding to the $\bar N$ measurements of the second step. Thus
\begin{equation}
    \label{eq:MLE}
    \hat\theta_{\textrm{MLE}}=\theta'\mp\frac{1}{2}\arccos\left(\frac{4s^2{\sum p^2_i}/{\bar
    N}-1-s^4}{1-s^4}\right)
\end{equation}
and the MLE is twice degenerate. There is, however a trivial way to
break this degeneracy by using the outcomes of the first step. The
prescription is to choose the solution closest to the rough
guess~$\hat\theta_0$. This estimator has an asymptotically vanishing
bias, whereas that corresponding to the other solution has a
constant bias which goes roughly as
\begin{equation}
    \E[\hat\theta]-\theta\simeq\arccos\left[\frac{4s^2\sigma^2(\theta',\theta)-1-s^4}{1-s^4}\right].
\end{equation}
Therefore, as the prior estimation gets accurate with increasing
$\bar N$, the separation of the two maxima of the likelihood
function remain constant. By choosing the maximum closest to
$\hat\theta_0$, one has an exponentially small error probability.

In summary, the optimal scheme goes as follows:
\begin{enumerate}
    \item Perform \textit{any} phase measurement (not necessarily optimal) on $N^\alpha$ ($1/2<\alpha<1$) copies of the signal state.
    From the outcomes, compute a preliminary estimation $\hat\theta_0$, which has a typical error $\delta\theta^2\simeq1/N^\alpha$.
    \item Measure the quadrature with $\theta'=\hat\theta_0+\Phi(s)/2$, where $\Phi(s)$ is given in Eq.~\eqref{eq:optimalPhi}, on the remaining $\bar N=N-N^\alpha$ copies.
    This corresponds to setting the phase reference (the Local Oscillator) to $\theta'$.
    \item The Maximum Likelihood Estimator is obtained by choosing
    the minus sign in Eq. \eqref{eq:MLE}.
\end{enumerate}
Note that one could equivalently write
$\theta'=\hat\theta_0-\Phi(s)/2$ and choose the plus sign in Eq.
\eqref{eq:MLE}. The variance of this estimator goes as
\begin{equation}
    \var[\hat\theta_{\textrm{MLE}}]\sim\frac{1}{F}=\frac{2s^4}{(1-s^4)^2}=\frac{1}{8(\langle n\rangle^2+\langle
    n\rangle)},
\end{equation}
and its bias vanishes asymptotically, therefore its MSE goes as
\begin{equation}
    \MSE[\hat\theta_{\textrm{MLE}}]\sim\frac{1}{8(\langle n\rangle^2+\langle
    n\rangle)},
\end{equation}
which is the Heisenberg limit presented in Sec.
\ref{sec:estimationth}.

\section{Conclusions}
We have seen that the phase-number Heisenberg inequality is valid
regardless of the existence of any phase operator, provided
$\langle\Delta\theta\rangle^2$ is regarded as a phase estimation
variance. It is just a consequence of the Cram\'er-Rao and
Braunstein-Caves inequalities.

We have seen that asymptotically the Heisenberg limit (optimal
estimation) can be attained by means of a (three-outcome) POVM
adaptive measurement. The physical implementation of such a
measurement is a demanding open problem. We have introduced an
\textit{approximate} MLE that provides an outstanding simplification
of the maximum likelihood analysis. Asymptotically this approximate
MLE and the exact one are shown to be equivalent.

 We have shown that the Heisenberg
limit can be also attained asymptotically by means of dyne
measurements and, surprisingly, only one quadrature needs to be
measured. This provides a remarkable economy of resources as
compared to POVM measurements such as an eight-port homodyne
detection with squeezing in the auxiliary port. The relative phase
between the Local Oscillator in homodyne detection and the signal
state has been computed and used to devise a two-step phase
estimation scheme which attains the Heisenberg limit asymptotically.

Remarkably enough, optimality is achieved with local measurements.
Hence, collective measurements prove of little use in the large
sample limit. This is a consequence of quantum estimation theory
applied to one-parameter problems.

Most of the schemes available so far succeed in attaining the
behavior $\var[\hat\theta]\sim\langle n\rangle^{-2}$ for $\langle
n\rangle$ large enough. The exact behavior is usually not considered
in the literature. This is, in our opinion, an important loophole in
phase estimation, since one is usually interested in low energy
states, where~$\langle n\rangle$ is small. An exception is the paper
by Yurke \emph{et al.} \cite{yurke86}, in which they introduce a
scheme whose variance comes close to our bounds for large $\langle
n\rangle$ \footnote{See eq. (10.39) in ref. \cite{yurke86}.}.
However, our scheme is asymptotically optimal and beats, to the
extent of our knowledge, any proposal presented up to this day.

\section{Acknowledgements}
The early stage of this research was carried out at the Max-Planck
Institut f\"ur Quantenoptik under the advice of Prof. I. Cirac.
Fruitful discussions with him, Dr. M. Wolf and their Theory Group at
the MPQ are gratefully acknowledged. Also discussions with E. Bagan,
R. Mu\~noz-Tapia, J. Calsamiglia and M. Mitchell have proved
fruitful. Finally, useful comments by the referee are acknowledged.
\appendix

\section{Averaging over the number of informative outcomes $\boldsymbol{N_{\textrm{inf}}}$}
\label{sec:appAverage} In this appendix we compute the average of
$\E_{\theta,N_{\textrm{inf}}}[(\hat\theta_{\textrm{MLE}}-\theta)^2]$
over the number of informative outcomes
$N_{\textrm{inf}}$, \emph{i.e.},
\begin{widetext}
\begin{eqnarray}
    \nonumber
    \E[(\hat\theta_{\textrm{MLE}}-\theta)^2]&=&\sum_{N_{\textrm{inf}}=0}^{\bar N}\E_{N_{\textrm{inf}}}[(\hat\theta_{\textrm{MLE}}-\theta)^2]{\binom{\bar N}{N_{\textrm{inf}}}}p^{N_{\textrm{inf}}}(1-p)^{\bar N-N_{\textrm{inf}}}\\
    &=&\frac{1}{4\langle\Delta n\rangle^2}\sum_{N_{\textrm{inf}}=1}^{\bar N}\frac{1}{N_{\textrm{inf}}}{\binom{\bar N}{N_{\textrm{inf}}}}p^{N_{\textrm{inf}}}(1-p)^{\bar
    N-N_{\textrm{inf}}}+\delta\theta^2 (1-p)^{\bar N}.
\end{eqnarray}
\end{widetext}
The first sum runs from $1$ to $\bar N$, thus taking into account
that the actual variance for the rare case $N_{\textrm{inf}}=0$ is
considered separately. We can compute the expectation value
$\langle{1}/{N_{\textrm{inf}}}\rangle$ by expanding
${1}/{N_{\textrm{inf}}}$ around the expectation value of
$N_{\textrm{inf}}$, which is $\langle N_{\textrm{inf}}\rangle=p\bar
N$, and keeping terms up to second order. We obtain
\begin{multline}
    \sum_{N_{\textrm{inf}}=1}^{\bar N}\frac{1}{N_{\textrm{inf}}}{\binom{\bar N}{N_{\textrm{inf}}}}p^{N_{\textrm{inf}}}(1-p)^{\bar N-N_{\textrm{inf}}}\\=\frac{1}{\bar Np}\left(1+\frac{1-p}{\bar Np}-(1-p)^{\bar
    N}\right).
\end{multline}
Inserting the value for $p=1+O(\delta\theta^2)$ yields
\begin{eqnarray}
    \E[(\hat\theta_{\textrm{MLE}}-\theta)^2]&=&\frac{1}{4\langle\Delta n\rangle^2\bar
    N}+O(\delta\theta^2).
\end{eqnarray}


\end{document}